\documentstyle[epsfig,psfig]{texas}

\begin{document}

\title{Similar Shot Noise in Cyg X-1, GRO J0422+32 and 1E 1724-3045}
\author{Wenfei Yu and Tipei Li}
\address{Laboratory for Cosmic Ray and High Energy Astrophysics\\
Institute of High Energy Physics\\
Chinese Academy of Sciences\\
P.O.Box: 918-3\\
Beijing, 100039\\
P. R. China\\
}
{\rm Email: yuwf@astrosv1.ihep.ac.cn, litp@astrosv1.ihep.ac.cn}

\begin{abstract}
We study the rapid variability in Cyg X-1 observed with 
the Rossi Timing Explorer (RXTE) on Jan 17 and 20 of 1997. 
The power spectra below 
200 Hz can be characterized by two ``shot noise'' components 
and a peaked-noise centered at 0.2 Hz. This is similar to those 
``shot noise'' observed in another black hole candidate 
GRO J0422+32 and a neutron star X-ray binary 1E 1724-3045. 
The similarity suggests that the generation mechanism of 
``shot noise'' and peaked-noise is probably similar in both 
black hole systems and neutron star systems, and in both 
high mass X-ray binaries (HMXBS) and low mass X-ray binaries 
(LMXBs). We have also analyzed $\sim$ 500 brightest shots 
selected from the Cyg X-1 light curves. The time scale of 
the spectral variation around the shot can be as long as 
$\sim$ a few seconds, consistent with Ginga results. 
The difference between the superposed shot profile in different 
energy bands can be attributed to the time lag and the 
different shot width between different 
energy bands. The shot width defined from the auto-correlation 
function of shot profile does not show a bimodal distribution.  
\end{abstract}

\section{Introduction}
The aperiodic X-ray variability of the black hole X-ray binary 
Cyg X-1 and other stellar black hole candidates has been described 
by the phenomenological ``shot noise'' models (e.g. \cite{ref1}\cite{ref2}
\cite{ref3}\cite{ref4} and references therein). 
Such rapid X-ray variability or flickering 
is more pronounced during their hard state
\cite{ref5}\cite{ref6}\cite{ref7}. In hard states, the 
power density spectra show a flat top followed by a power law at a 
certain break frequency. The higher amplitude of the rapid variability is, 
the lower the break frequency is. This suggests that the shot properties 
vary with time in the hard states. 

In the framework of ``shot noise'' models, the properties of the shots 
and superposed shot profiles in Cyg X-1 have been studied with data 
obtained from Uhuru, HEAO 1 A-2, EXOSAT/ME, Ginga, and RXTE/PCA (see 
\cite{ref2} and references therein;\cite{ref8}\cite{ref9}\cite{ref10}). 
In summary, the shot properties are as follows: 
(1) The shot has nearly time symmetric rise and decay lasting for up 
to a few seconds; (2) The energy spectrum of the shot changes with time. 
(3) The shot duration changes with the states.  

Recently, striking timing similarities of the power density spectra 
between the black hole candidate GRO J0422+32 (Nova Persei 1992)
\cite{ref11}\cite{ref12} and the X-ray burster 1E 1724-3045\cite{ref13}
 were reported. 
In this paper, we show similar aperiodic variability in Cyg X-1 
in hard state and compare the shot noise properties with those of 
GRO J0422+32 and 1E 1724-3045. We also present our results from 
the study of about $\sim$ 500 shots observed in Cyg X-1 with RXTE/PCA. 

\section{RXTE/PCA Observation of Cyg X-1}
We analyze the data obtained from RXTE/PCA observations conducted on 
1997 Jan 17 and 20. The entire PCA observation lasted for about 
12.5 hours. The average count rates in the entire PCA band for the two 
days are 4330 and 3880 cps, respectively. High time resolution 
($\sim$ 2$^{-12}$ s) {\it Single-Bit} mode 
data in the energy range 1.0-5.1 keV, 5.1-8.7 keV, 8.7-18.3 keV 
and 18.3-98.5 keV were used in studying the aperiodic variability. 
We combine the data in the 4 bands to calculate the average power 
density spectra (PDS). The average PDS of Jan 20 is plotted in Fig.1. 
The PDS displays a flat top below a low frequency break around 0.03 Hz, 
a peaked-noise component centered at 0.2 Hz with a FWHM $\sim$ 0.2 Hz 
and a second break frequency around 3 Hz. 
\begin{figure}
\centering
\psfig{file=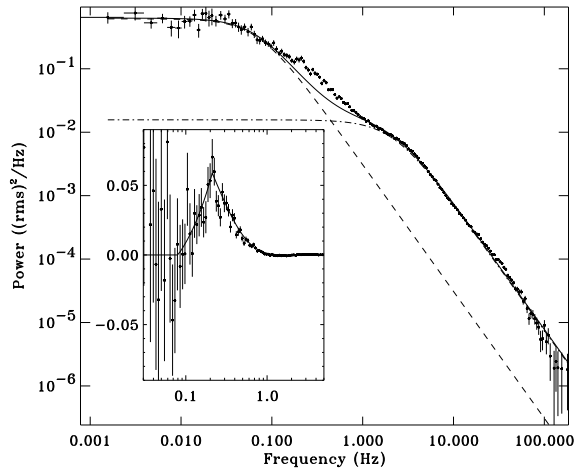,width=8cm}
\caption{Average power density spectra of Cyg X-1 as observed 
with RXTE/PCA on 1997 Jan 20. Two band-limit Lorentzian functions were 
fit to the PDS below 0.1 Hz and above 1 Hz (dashed line and dash-dot line). 
The residuals are shown in 
the inset panel, indicating a peaked noise component centered at 0.21 Hz. 
A model of a linear rise followed by an exponential decay was fit to the 
residuals in the inset panel (solid line).}\label{fig1}
\end{figure}

Similar to the previous study of GRO J0422+32\cite{ref11}\cite{ref12} 
and 1E 1724-3045\cite{ref13}, we apply a model consisting of 
two ``shot noise'' components characterized by two Lorentzian functions 
in the frequency range below 0.1 Hz and above 1.0 Hz (dashed line and
dash-dot line), and fit the residual 
noise power in the frequency range 0.1--1.0 Hz with a model 
composed of a linear rise with an exponential decay (solid line in the 
inset panel). The model is shown as solid line in the Figure 1.

\section{Comparison with Observations of GRO J0422+32 and 1E 1724-3045}
The PDS of OSSE observation to GRO J0422+32\cite{ref11}\cite{ref12} 
and the PDS of 1E 1724-3045 observed with RXTE/PCA\cite{ref13} 
can also be characterized by two ``shot noise'' components and one 
peaked-noise feature. The characteristic duration of the two ``shot 
noise'' components ($\tau_1$ or $\tau_2$) 
is represented by the Half Width of Half Maximum 
(HWHM) of each Lorentzian as 
$$\tau_{1,2}=\frac{1}{2\pi~HWHM}$$ 
The parameters of the noise components of the three sources are 
compared in Table 1.

\section{X-ray Shots of Cyg X-1 in Hard State}
To investigate the above interpretations in terms of ``shot noise'', 
we study the X-ray shots observed with RXTE/PCA in the energy range 
$\sim$ 2--60 keV. The X-ray Shots were selected in the light curve. 
The criteria are that their peak X-ray counts in 0.125 s bin should be 
larger than 1250, and should be the maximum within the neighboring 
5.0 s on both sides. In the entire observation, we have found 513 
shots which meet the requirement.

\begin{figure}
\centering
\psfig{file=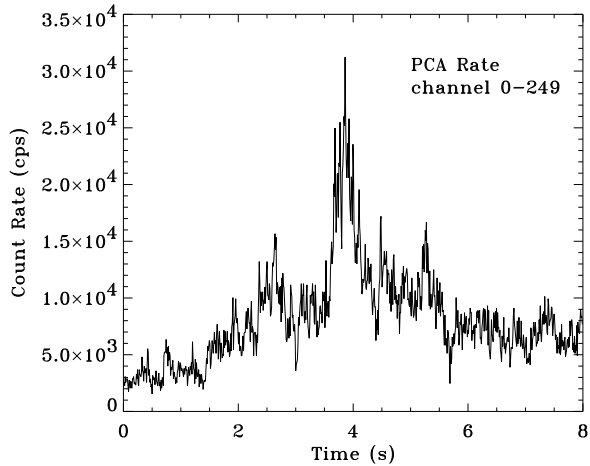,width=8cm}
\caption{Light curve of one of the brightest shot in the entire 
PCA energy range.}\label{fig2}
\end{figure}
 
\paragraph{Shot Width} 
The width of each shot is derived from the auto-correlation 
coefficients of 10 s light curve around each shot, $A(i)$. They are 
defined as follows:
$$A(i)=\frac{\sum_{k=0}^{N-i-1}(X_{k}-\bar{X})(X_{k+i}-\bar{X})}
{\sum_{k=0}^{N-1}(X_{k}-\bar{X})^2}$$
where $\delta{t}$ is $\sim$ 2.44 ms, the time resolution of the light 
curves, and $A(i)$ ($i=0,...,N-1$) the 
auto-correlation coefficient at $i\delta{t}$. Then we 
define the average shot width as 
 $${T_{width}}=2.0\times\sqrt{\frac{\sum_{k=1}^{M}k^2A(k)+(0.25)^2A(0)}{\sum_{k=0}^{M}A(k)}}\delta{t}$$
where 0.25 represents the average time shift of the central bin A(0), and 
$M$ is the maximum of $i$ with $A(i)$ no less than 0.0 in the main 
peak of the auto-correlation function.
In Fig.2, we show the profile of one of the brightest shot. The profile was 
obtained from the {\it Standard 1} data mode. The corresponding 
auto-correlation coefficients are 
shown in Fig.3. They were obtained from {\it Single-Bit } data and bined 
to a time resolution of $\sim$ 2.44 ms. The coefficients used in the 
calculation of shot width are shaded in the figure. The distribution of 
shot width for the 513 shots is shown in Fig.4. 
\begin{figure}
\centering
\psfig{file=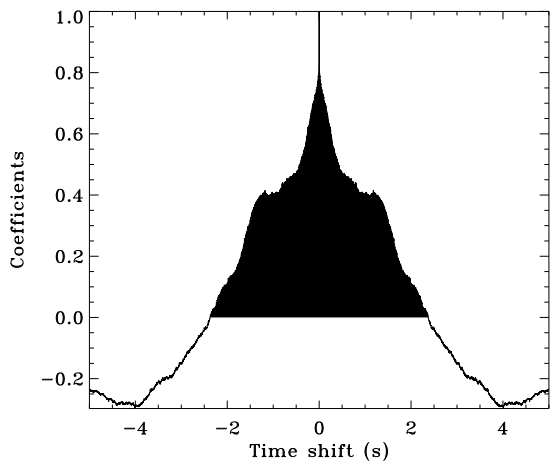,width=8cm}
\caption{The auto-correlation coefficients of the shot 
shown in Fig.2. Shaded region corresponds to the data 
used in calculating the shot width.}\label{fig3}
\end{figure}

\begin{figure}
\centering
\psfig{file=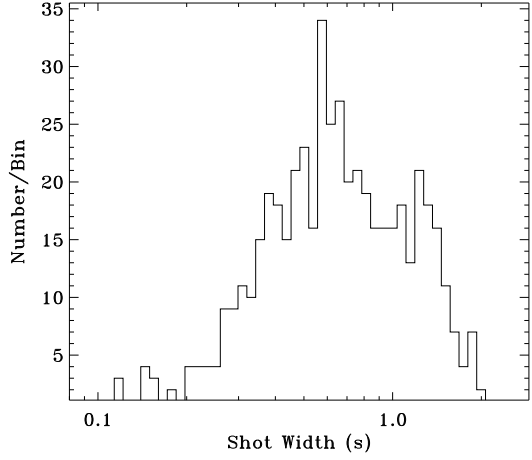,width=8cm}
\caption{The distribution of the shot width.}\label{fig4}
\end{figure}

\paragraph{Peak-aligned Shot Profile and Spectral Variation}		  
We superposed the 513 shots by aligning them at the center of their 0.125 s 
peak bins (combine all 4 energy channels). The time resolution used in the 
alignment is $\sim$ 2.44 ms. In the upper panel of Fig.5, we show the 
peak-aligned shot profiles of the soft band (channel 1+2) and the hard band 
(channel 3+4), respectively. They were normalized to their peak counts. In 
the lower panel of the same figure, we plot the residuals of the 
profile subtraction (3+4)--(1+2). In general, the peak-aligned profiles in 
Fig.5 shows: (1)The time scale of the shot rise in the hard band is smaller 
than that in the soft band; (2)The shot decay in the hard band is slower than 
the decay in the soft band; (3)There are more hard photons after the shot peak 
than those before the shot peak, indicating a spectral hardening.
\begin{figure}
\centering
\psfig{file=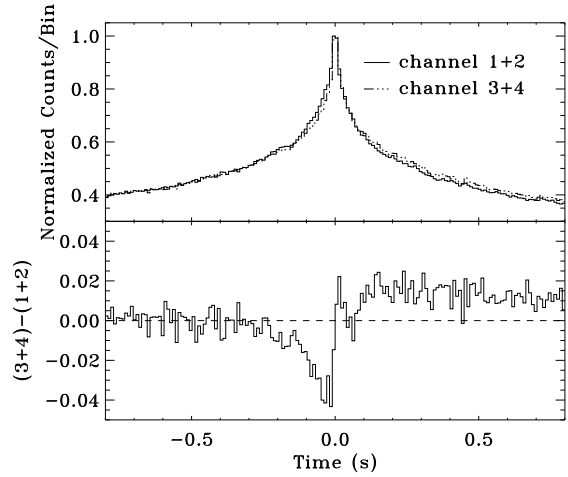,width=8cm}
\caption{The peak-aligned shot profile in the soft band (1+2) and 
the hard band (3+4)}\label{fig5}
\end{figure}

Three factors would introduce the difference between the profiles in the soft 
band and those in the hard band and the observed residuals in the lower panel 
of Fig.5. One is the time lag of the shot rise between the soft band and the 
hard band. The other is the slower decay in the hard band compared with that 
of the soft band. The third is that the shot peak in the hard band is narrower 
than that in the soft band. To study the spectral variation, we plot the 
ratio between the peak-aligned profile in the hard band and that in the 
soft band in 
Fig.6. Both sides around the peak were fit to a 5-degree polynomial. 
The spectra after the shot peak is harder than that before the shot 
peak, as shown in Fig.6. This is 
consistent with Ginga results\cite{ref8}.   
\begin{figure}
\centering
\psfig{file=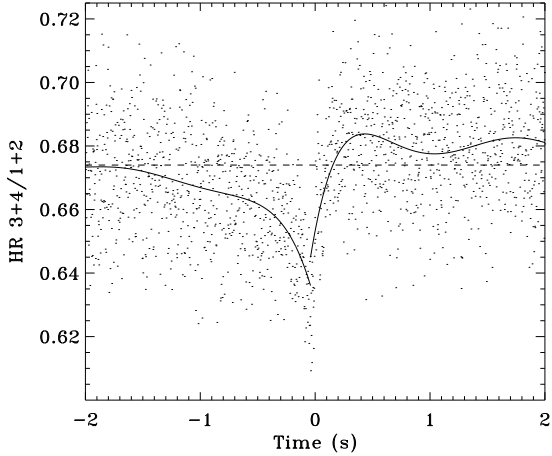,width=8cm}
\caption{The ratio between the profiles in the two bands 
(1+2) and (3+4). Two 5-degree polynomials were fit to the data before 
and after the shot peak. They suggest that the spectra during the 
shot decay are harder.}\label{fig6}
\end{figure}

\section{Summary}

In summary, we have obtained the following results:
\begin{itemize}
\item We have found that the aperiodic X-ray variability in Cyg X-1 in 
hard state is similar to those observed in another black hole candidate 
GRO J0422+32 and a neutron star X-ray binary 1E 1724-3045. Based on the 
mass of companion star and the mass of the central object in the three 
X-ray binaries, we conclude that the generation mechanism of the X-ray 
shots is probably independent on the mass of the companion star, the mass of 
the central compact object (BH or NS), and the type of accretion.  
\item The spectral evolution around the shot peak could last for as 
long as a few seconds, and there is a spectral hardening after the 
shot peak. These are consistent with the previous study of the X-ray 
shots in Cyg X-1 observed with Ginga\cite{ref8}. 
\item The duration of $\sim$ 513 bright shots in Cyg X-1, 
defined from their auto-correlation coefficients, ranging from 
$\sim$ 0.1 s to $\sim$ 2.0 s, is not bimodally distributed. This does 
not support the assumption that there are two kinds of shots with different 
duration as inferred from the power spectra. Thus the 
attribution of each 
of the noise component to a group of shots with a certain characteristic 
duration is probably wrong.  
\end{itemize}

\begin{table}
\begin{center}
\small
{Table 1. A Comparison of GRO J0422+32$^{(1)}$, 1E 1724-3045$^{(2)}$ and Cyg
X-1$^{(3)}$}

\begin{tabular}{|c|c|c|c|}
\hline
Source Name & GRO J0422+32 & 1E 1724-3045 & Cyg X-1 \\ \hline
Binary Type & LMXB & LMXB & HMXB \\ \hline
Accretion Type& Disk & Disk & Wind + Disk \\ \hline
Compact Object & BH & NS & BH \\  \hline
Central Mass (M$\odot$)$^{(4)}$ & 3.2-3.9 or $>$ 9 & 1.4-2.0 & 16$\pm$5 \\ \hline
Source State & Hard & Island, Low & Hard \\ \hline
Rms Amplitude & 30\% (35-60 keV), 40\% (75-175 keV)& 25\% (2-20 keV) & 39\% (1-98.5 keV)\\ \hline
Characteristic Duration $\tau_{1}$ (s) & $\sim$ 2.2& $\sim$ 0.7 & $\sim$ 2.2 \\ \hline
Characteristic Duration $\tau_{2}$ (s) & $\sim$ 0.050 & $\sim$ 0.017 & $\sim$ 0.070 \\ \hline
Peaked-Noise Frequency (Hz) & $\sim$ 0.2 & $\sim$ 0.8 & $\sim$ 0.2 \\ \hline
\end{tabular}\\
\vspace{0.5cm}
\end{center}
{\it Note:} \\
(1) \cite{ref11}; (2) \cite{ref13}; 
(3) Yu, W. 1998, Ph.D thesis; (4) \cite{ref14} and references therein.
\end{table}

\section*{Acknowledgments}
WY appreciate various supports and helpful discussions and comments by 
Prof. J. Van Paradijs, Dr. C. Kouveliotou and Dr. M. Finger at NASA/MSFC.

\section*{References}
%%% Standard "thebibliography" LaTeX environment. References sorted BY
%%% ORDER OF APPEARANCE for hypertext links and cross-referencing.
%%%

\end{document}